# Picosecond two-photon absorption laser induced fluorescence (ps-TALIF) in krypton: the role of photoionization on the density depletion of the fluorescing state Kr 5p´[3/2]$_2$

K. Gazeli, X. Aubert, S. Prasanna, C. Y. Duluard, G. Lombardi, K. Hassouni

University Sorbonne Paris Nord, LSPM, CNRS, UPR 3407, F-93430 Villetaneuse, France

Corresponding authors : kristaq.gazeli@lspm.cnrs.fr ; xavier.aubert@lspm.cnrs.fr ; guillaume.lombardi@lspm.cnrs.fr

**Abstract**

The present study focuses on the application of a picosecond (ps) TALIF technique in krypton (Kr) at variable pressure (0.1–10 mbar). The laser intensity ($I$, units W.cm$^{-2}$) is tuned between 1 and 480 MW.cm$^{-2}$, and the depletion of the density of the Kr 5p´[3/2]$_2$ fluorescing state through photoionization (PIN) and amplified stimulated emission (ASE) is investigated. This is done by combining TALIF experiments with a simple 0D numerical model. We demonstrate that for a gas pressure of 3 mbar and $15 < I \leq 480$ MW.cm$^{-2}$, a saturated fluorescence signal is obtained, which is largely attributed to PIN, ASE being negligible. Also, a broadening of the two-photon absorption line (i.e. 4p$^6$ $^1$S$_0$→→5p´[3/2]$_2$) is recorded due to the production of charged species through PIN, inducing a Stark effect. For $I \leq 15$ MW.cm$^{-2}$, though, PIN is significantly limited, the absorption line is noticeably narrowed, and the quadratic dependence of the TALIF signal intensity versus the laser energy is obtained. Thus, in this case, the investigated Kr TALIF scheme, using the 5p´[3/2]$_2$→5s[3/2]$_1$ fluorescence channel, can be used for calibration purposes in ps-TALIF experiments. These results are of interest for fundamental research since most ps-TALIF studies performed in Kr do not investigate in detail the role of PIN and ASE on the depletion of the Kr 5p´[3/2]$_2$ state density. Moreover, this work contributes to the development of ps-TALIF for determining absolute densities and quenching coefficients of H and N atoms in plasmas. This is useful in numerous plasma-based applications (e.g. thin film synthesis, biomedical treatments, plasma-assisted combustion, …), for which the knowledge of the density/kinetics of reactive atoms is essential.

**Keywords:** picosecond, TALIF, ultrafast laser, laser diagnostics, krypton.

## 1. Introduction

The implementation of fast (nanosecond–ns) and ultrafast (picosecond–ps and femtosecond–fs) optical diagnostics for the study of ionized gases is nowadays among the principal research axes of different research groups [1-16]. Such diagnostics aim for the precise identification and fine tuning of essential physical quantities, leading to a better understanding of transient phenomena in reactive plasmas. Typical examples refer to the use of (i) two-photon absorption laser induced fluorescence (TALIF) to obtain atomic densities [1,5,7–14], (ii) electric-field-induced second harmonic generation to infer electric field dynamics [1,6,10], and (iii) Thomson scattering to access electron density and temperature [15,16]. These measurements can help in the development of diverse plasma-based applications, e.g. synthesis of advanced nanomaterials [17,18], biomedical [19,20], public security [21], and environmental applications [22,23], where the above plasma quantities can play a key role.

Particularly, TALIF is a reliable technique that has been extensively applied over the last decades to probe reactive atoms (mostly H, N, and O) in collisional media [1]. TALIF offers high sensitivity with a very good spatial and temporal resolution. The latter is defined by the duration of the laser pulse, which can be either in the ns [7–10,24,25], ps [14,26–28] or fs [11–13,29] time-scale. In general, ns-TALIF is consistent for measurements at gas pressures lower than 10 mbar [30]. However, with increasing gas pressure, the collisional quenching can become the major loss mechanism of excited states of interest depending on the species and the quencher nature [7]. For instance, the quenching time of laser-excited H atoms in flames falls well below ns at atmospheric pressure [27], which is shorter than the pulse duration of ns lasers (typically few ns [7–10,24,25]). In this case, to infer the quenching rate of species from the fluorescence signal, the laser pulse duration must be much shorter than the TALIF decay time. Otherwise, more complex model-based signal corrections need to be performed [8,31]. For accurate measurements at pressures close to the atmospheric (or even higher [13]), a better approach would be to use ultrafast lasers. These can deliver (i) sufficient peak intensities ($I$, units W.cm$^{-2}$) to achieve the necessary two-photon transitions, and (ii) pulse widths down to the ps and fs timescales.

The use of ultrafast lasers for probing reactive species in flames has been already reported in the past decades in excellent works of different groups [1,26–28,32]. However, the implementation of mostly ps– [14], and fs–TALIF [11–13,29] for quench-free measurements in collisional plasmas is very recent. In ps/fs TALIF, the relatively high peak intensities of the laser pulses used can lead to saturated fluorescence regimes [5]. This is due to a substantial depletion of the densities of the laser-excited states through photoionization (PIN) and/or amplified stimulated emission (ASE) [1,5,26–28,33]. These processes must be studied in detail so as to be carefully avoided when absolute density measurements are sought from TALIF signals.

Such an understanding is particularly needed in the case of ps-TALIF in krypton. This rare gas has been well-utilized as a calibrating species for the measurement of H- and N-atom densities with ns-TALIF [7–10,25,30]. Non-saturating regimes with a quadratic dependence between the TALIF signal intensity and the laser energy ($E_{Laser}$) were reported [8,10,25,30]. Within these regimes, the kinetic of the fluorescing state is dominated by the processes involved in the TALIF scheme, and the fluorescence signal intensity is proportional to the ground-state Kr atom density. However, the use of Kr gas in ps-TALIF for studies of reactive plasmas is very recent [14,34], and requires further assessment of the conditions where Kr atom may be still considered for calibration purposes.

Based on the above, the present study focuses on the fundamental investigation of the relative predominance of the processes involved in the ps-TALIF scheme of Kr atom. In particular, the critical role of PIN on the depletion of the Kr 5p´[3/2]$_2$ fluorescing state at high $E_{Laser}$, as well as conditions leading to a minor PIN effect are demonstrated. The later also allow for the quadratic dependence on $E_{Laser}$ of the TALIF-signal intensity to be ensured. Thus, besides being a fundamental investigation, our work contributes to the development of ps-TALIF in Kr gas to calibrate measurements of H- and N-atom densities in reactive plasmas.

## 2. Experimental setup

**Fig. 1** illustrates the experimental setup used to perform ps-TALIF studies in krypton gas (Air liquide, 99.998% purity). The gas was introduced in a low-pressure gas cell, which offered optical accesses through two fused silica (for the laser) and one BK7 (for the TALIF) windows, suitable for TALIF diagnostics in the UV–NIR range. Besides, it allowed for the precise control ($\pm 5\%$) of the Kr gas pressure ($P_{Kr}$) between 0.1 and 10 mbar using a MKS Baratron® capacitance manometer (pressure range: 0.01–10 mbar).

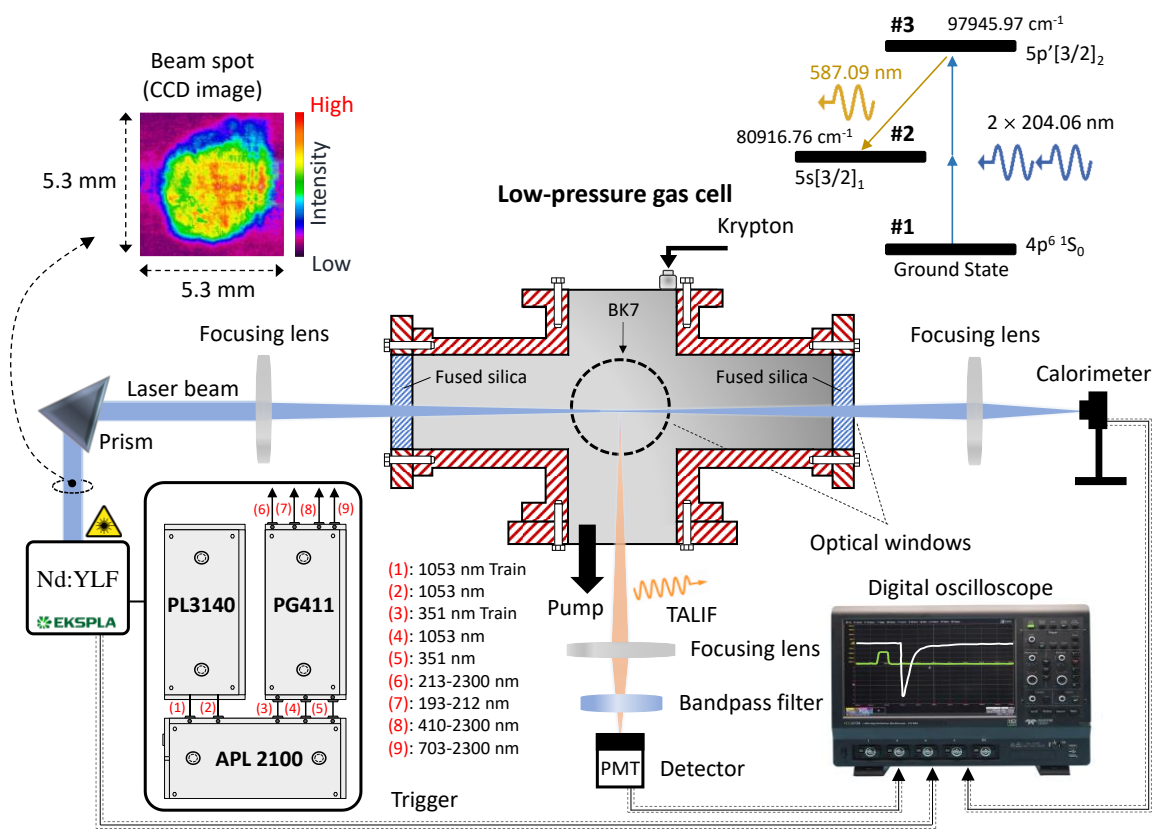

**Figure 1.** Experimental setup of ps-TALIF in Kr gas. The two-photon excitation scheme and the investigated fluorescence transition are given at the top right. The laser beam at the output (7) of PG411 (EKSPLA®) is fairly homogeneous as shown at the top left (measured with a CCD camera beam-profiler).

In the following parts of the paper, one will note the ground-state of Kr ($4p^6\ {}^1S_0$) as level #1, the laser-excited state Kr $5p'[3/2]_2$ as level #3, and the de-excitation state Kr $5s[3/2]_1$ as level #2. To induce a two-photon excitation from level #1 to level #3, and analyze the subsequent fluorescence decay to level #2 (i.e. #3→#2 transition at λ=587.09 nm, as that used in [9], see **Fig. 1**), an ultrafast laser from EKSPLA® was used. It was composed of three units: i) a Nd:YLF pump laser (PL3140, 50 mJ max output, ∼10 ps pulse width) selectively producing a fundamental output of 1053 nm in two modes: pulse train or 5 Hz repetition rate, ii) a harmonic generator (APL2100, 50 mJ max output, <15 ps pulse width) for fundamental frequency doubling (527 nm) and tripling (351 nm), and iii) a solid state optical parametric generator (PG411, 2 mJ max output, 10 ps pulse width, ∼4 cm$^{-1}$ linewidth as it is assessed by the manufacturer) generating laser pulses between 193 and 2300 nm at different tuning ranges ((6)–(9)). For the present study, the laser pulse repetition rate was kept at 5 Hz (output (2) of PL3140). The suitable experimental wavelength for the two-photon excitation of ground-state Kr in our case was $\lambda_{c_{Kr}}$=204.06 nm ($\tilde{v}_{c_{Kr}}=1/\lambda_{c_{Kr}}$=49005.19 cm$^{-1}$) and was tuned within the range 193–212 nm of the output (7) of PG411. The laser spot after the output (7) of PG411 was fairly homogeneous with a diameter of less than 5 mm (see **Fig. 1**). Using a series of prisms, the laser light was guided to a lens (35 cm focal length) focusing it at the center of the gas cell. Then it was collected with a second identical lens, as shown in **Fig. 1**. A calorimeter (Coherent J-10MB-LE, 11630 V/J) was placed after the second lens to measure the laser energy per pulse. Its output was recorded with a high-definition oscilloscope (Lecroy HDO6104, 1 GHz, 2.5 Gs/s). TALIF was collected at 90º angle with respect to the laser beam by using a third focusing lens, and a bandpass filter (Semrock FF02-586/15-25, center wavelength: 586 nm, bandwidth: 20.5 nm) to eliminate any stray light. Then, it was focused to the entrance slit (500 μm width) of a fast photomultiplier tube (PMT, Hamamatsu R9110) and stored in the same oscilloscope. The features of the TALIF signal were studied for $E_{Laser}$ values ranging between 0.1 and 13 μJ/pulse, corresponding to $I$=4 and 52 MW.cm$^{-2}$ ($I=E_{Laser}/(D \times \tau_L)$), respectively, for a beam diameter ($D$) of 500±100 μm at the collection point, and $\tau_L$=10 ps (laser pulse duration of PG411 indicated by the manufacturer). The variation of $E_{Laser}$ was achieved by combining different neutral density filters placed after the output (7) of PG411. The shot-to-shot energy fluctuation was <10% (measured over 1000 laser pulses during different days, the room's temperature being controlled *via* air conditioning).

## 3. Results and discussion

The spectrally– and temporally–integrated TALIF signal ($\iint S(\tilde{v},t)d\tilde{v}dt=S_{\tilde{v}t}$) depends on the PMT quantum efficiency ($\eta$), the transmission of the optics used ($T$), the solid angle of detection (ΔΩ), the fluorescence yield ($a_{32} = A_{32}/(A_3 + Q_3)$), where $A_{32}$ is the Einstein coefficient for the fluorescence channel #3→#2, $A_3$ is the sum of Einstein coefficients for all radiative channels from level #3, and $Q_3$ is the quenching rate of level #3, the Kr-atom density ($n_{Kr}$), the photon statistic factor ($G^{(2)}$), the two-photon absorption cross-section ($\sigma^{(2)}$), the time-integrated squared laser intensity ($\int I^2 dt \propto E_{Laser}^2$), the squared laser photon energy ($(h\nu)^2$), and the observed excitation volume ($V$), as follows [7]:

$$S_{\tilde{\nu}t} = \eta T \frac{\Delta\Omega}{4\pi} a_{32} n_{Kr} G^{(2)} \frac{\sigma^{(2)}}{(h\nu)^2} V \int I^2 dt \rightarrow S_{\tilde{\nu}t} \propto E_{Laser}^2 \qquad (1)$$

This is experimentally evaluated by capturing TALIF signals at different laser frequencies around the peak two-photon excitation frequency, followed by temporal and spectral integration. In the present work, the spectral profile of the two-photon absorption line was acquired over a laser-wavelength range of 250 pm with a step of 2 pm (see later in the text). Based on eq. (1), for reliable density measurements, $S_{\tilde{\nu}t}$ should exhibit a quadratic dependence on $E_{Laser}$ (quadratic regime) [7]. Thus, a linear regression of the experimental data on $S_{\tilde{\nu}t}$ versus $E_{Laser}^2$ should give a slope of 1. This is plotted in **Fig. 2**. For $E_{Laser}<0.45$ µJ/pulse (i.e. $I<18$ MW.cm$^{-2}$, green zone), a slope of 0.97±0.04 is obtained (red line), validating the quadratic regime. However, for $I \geq 18$ MW.cm$^{-2}$, a slope of 0.5±0.01 is found (blue line within the yellow zone). This implies that other physical processes are involved in the depletion of the fluorescing state, namely PIN and possibly ASE [5,27]. Thus, eq. (1) is not valid for $I \geq 18$ MW.cm$^{-2}$.

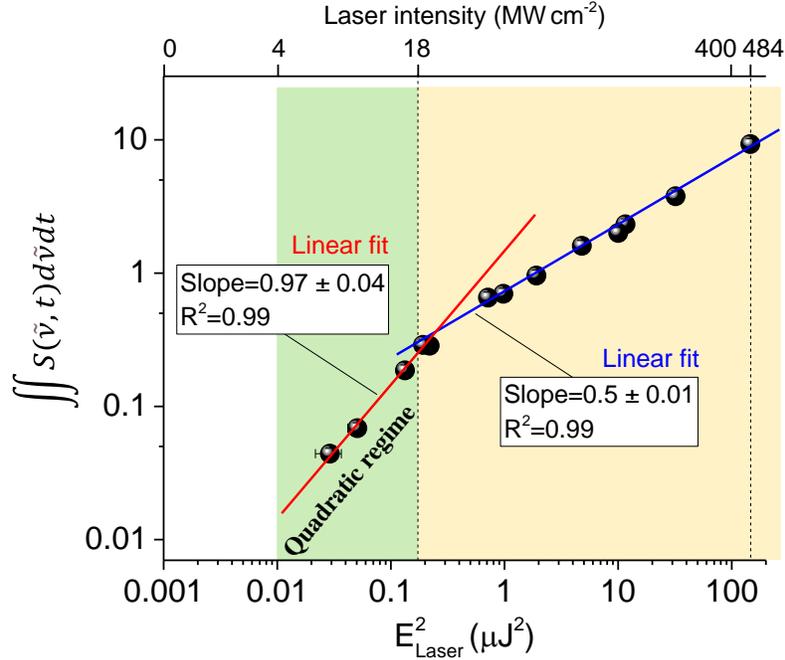

**Figure 2.** $S_{\tilde{\nu}t} = \iint S(\tilde{\nu},t)d\tilde{\nu}dt$ versus $E_{Laser}^2$ for $I<500$ MW.cm$^{-2}$ (top) and $E_{Laser}=0.1–13$ µJ/pulse (bottom) at $P_{Kr}=3$ mbar. The green and yellow zones indicate quadratic and saturation regimes, respectively.

When using ultrafast TALIF to probe atomic densities, the validity of eq. (1) (quadratic regime) is ensured if two conditions are satisfied [5]: (*c1*) the depletion of an atomic ground state (*i*) by the laser is negligible; to achieve that, $\tau_L$ must be much lower than the characteristic time of the two-photon absorption process, $1/W_{ij}$, where $W_{ij}$ is the frequency of the two-photon excitation from the ground to an excited state (*j*). This translates to $W_{ij} \times \tau_L \ll 1$. (*c2*) the depletion rate of the laser-excited state (level #3) through PIN and ASE must be much lower than $A_3$ and $Q_3$.

In **Fig. 2**, the above two conditions should be valid for $I<18$ MW.cm$^{-2}$, where the quadratic regime is obtained (green zone). Similarly, in [14,34], the quadratic regime was clearly achieved when using ps-TALIF (30 ps laser pulses) in Kr gas. However, in those studies, Kr gas was mostly used to calibrate absolute density measurements of H atoms in plasma jets. Concerning Kr atoms, they were laser-excited at $E_{Laser}$=0.28 µJ/pulse, which was well-below the TALIF saturation limit. Thus, for a laser-beam diameter of ~250 µm (given in [34]) at the TALIF collection point, $I_{[14,34]} \cong 8.2$ MW.cm$^{-2}$, which is consistent with our finding above. Our study here aims further to analyze the processes leading to the depletion of the fluorescing state while using 3 times shorter laser pulses (10 ps). Quadratic regimes have been also achieved when using fs–TALIF in Kr gas by constructing laser sheets so as to significantly decrease $I$ [12,35]. Therefore, our work and the excellent studies [12,14,34,35], contribute to the development of ultrafast TALIF diagnostics for determining absolute densities and quenching coefficients of H and N atoms in reactive plasmas.

For $I \geq 18$ MW.cm$^{-2}$ in **Fig. 2** (yellow zone), a lower slope (~0.5) of the linear regression is revealed, indicating a non-quadratic dependence on $E_{Laser}$ of the TALIF signal intensity. To understand this behavior, the validity of the above conditions (*c1*) and (*c2*) must be checked. For condition (*c1*) we have $W_{ij} = \frac{\sigma_{Kr}^{(2)} g(v_0) G^{(2)} I^2}{(hc)^2 \tilde{v}_{c_{Kr}}^2}$ [5] and $\tau_L$=10 ps. The value of $\sigma_{Kr}^{(2)}$, inferred from $\sigma_{Kr}^{(2)}/\sigma_H^{(2)}$=0.62 [7] (with 50% uncertainty) and $\sigma_H^{(2)}$= 1.77 ×10$^{-35}$ cm$^4$ [5,36], is of 1.09×10$^{-35}$ cm$^4$. For $I_{max}$=484 MW.cm$^{-2}$ ($E_{Laser}$=12.1 µJ/pulse), i.e. the highest laser intensity in **Fig. 2**, and $g(v_0)G^{(2)} \approx \frac{0.94}{\sim 4 \text{ cm}^{-1}} \times 2 \approx 1.56 \times 10^{-11}$ s (considering a Gaussian absorption-line profile dominated by the laser linewidth [5]), $W_{ij} \cong 4.19 \times 10^7$ s$^{-1}$ and, thus, $W_{ij} \times \tau_L \cong 4.19 \times 10^{-4} \ll 1$, satisfying the condition (*c1*) within the yellow zone. Thus, the condition (*c2*) regarding PIN and ASE should not be valid in the yellow zone, which is checked below by analyzing the time-decay of fluorescence signals.

The impact of $P_{Kr}$ and $E_{Laser}$ on the experimentally-obtained PMT signal $S_t = S(\tilde{v} = \tilde{v}_{c_{Kr}}, t)$ is shown in **Fig. 3**. The solid red and black lines refer to an $S_t$ recorded using a 586 nm-centred interference filter for the corresponding $E_{Laser}$ values: i) 0.4 µJ/pulse, i.e. within the quadratic regime (at $P_{Kr}$=3 mbar), and ii) 12.1 µJ/pulse, which is the upper $E_{Laser}$ limit studied outside the quadratic regime (see **Fig. 2**). For $E_{Laser} \leq 0.4$ µJ/pulse and $P_{Kr}$ up to 10 mbar in **Fig. 3**, the decaying part of the TALIF (solid red) follows an exponential variation with a time-constant $\tau$. Then, the total decay rates of the excited state ($1/\tau$) are in fact the sum of the natural (radiative) decay rate ($1/\tau_{Kr*}$) and the collisional quenching rate of Kr 5p´[3/2]$_2$ such that

$$\frac{1}{\tau} = \frac{1}{\tau_{Kr*}} + k_{Kr*} n_{Kr} \qquad (2)$$

Where $k_{Kr*}$ and $n_{Kr}$ are the collisional quenching rate and the Kr atom density, respectively [7]. $\tau_{Kr*}$ was evaluated for $E_{Laser} \leq 0.4$ µJ/pulse, and $P_{Kr}$=0.1–0.2 mbar, for which the collisional quenching was negligible. It was found to be of 34.6±2.3 ns, which is consistent with the values of 34.1 ns, 35.4±2.7 ns, and 33.6±1.1 ns reported in [7,12,37]. Further, from the Stern-Volmer plot [7], i.e. plotting of $\frac{1}{\tau}$ versus $n_{Kr}$ in the pressure range 0.1–10 mbar, $k_{Kr*}$ was found to be of 1.7±0.2×10$^{-10}$ cm$^3$ s$^{-1}$, which is in the range of values reported in the literature [7,12,37,38].

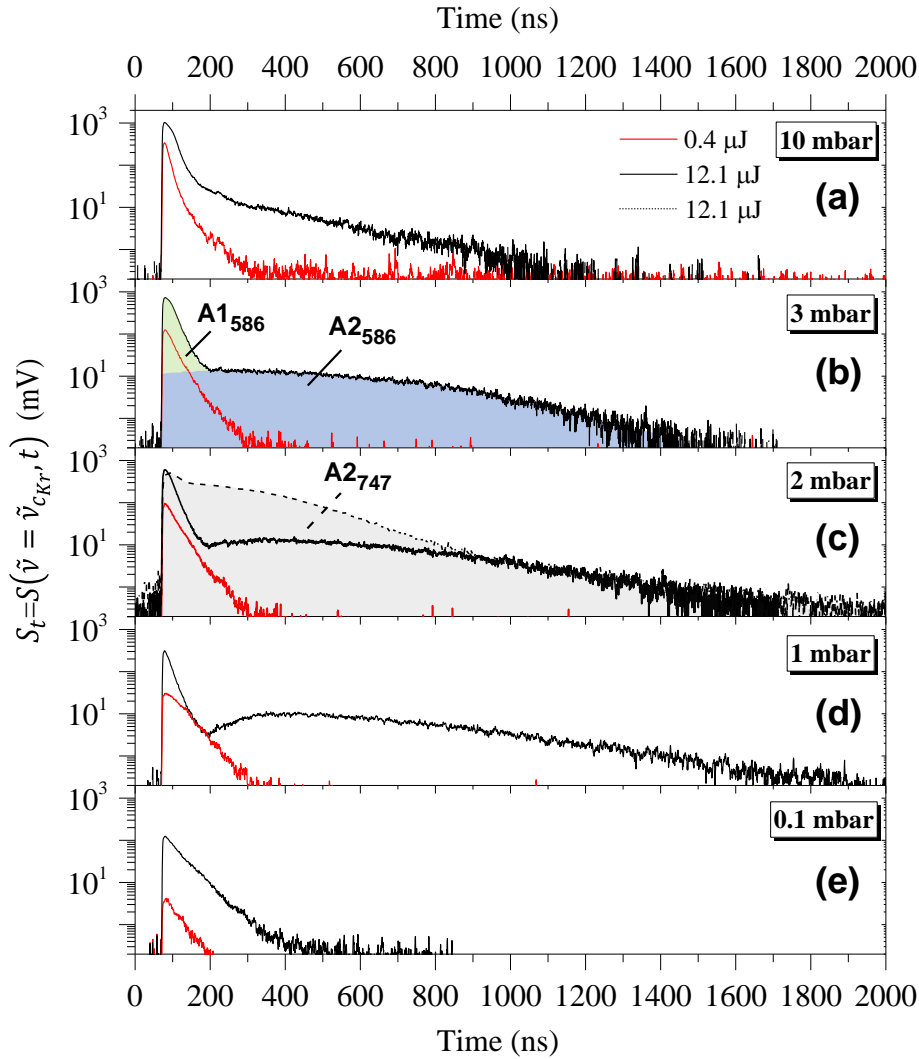

**Figure 3**. PMT signals $S_t=S(\tilde{v}=\tilde{v}_{c_{Kr}},t)$ for $P_{Kr}$=0.1–10 mbar at low (0.4 μJ/pulse – solid red) and high (12.1 μJ/pulse – solid black) $E_{Laser}$, recorded with a 586 nm-centred interference filter. Black solid signals in (a)-(d), are saturated consisting of two areas: $A1_{586}$ and $A2_{586}$ (indicated in (b) for $P_{Kr}$=3 mbar). In (c), the black dashed signal, which is essentially defined by the grey area $A2_{747}$ ($A1_{747}$ being very weak), is recorded when using a 747 nm-centred interference filter.

Interestingly, for temporal signals in the saturated regime ($E_{Laser}$=12.1 μJ/pulse and $P_{Kr}$>0.1 mbar in **Fig. 3**), two distinct signal-decay areas are revealed: $A1_{586}$ and $A2_{586}$, as indicated in **Fig. 3b**. The transition from $A1_{586}$ to $A2_{586}$ is observed at around 200-300 ns almost independently of $P_{Kr}$. $A1_{586}$ has a short duration and corresponds to the decay of Kr 5p´[3/2]$_2$ laser-excited state, which is characterized by an exponential function with a time constant $\tau$. On the other hand, $A2_{586}$, which forms almost immediately after the laser pulse (see below for justification), and is responsible for the inflection point at around 200-300 ns, has a larger duration depending on the gas pressure. This suggests that there could be additional mechanisms leading to the formation of different excited states of Kr atom. In fact, the 586 nm-centred interference filter used in this work could capture only three Kr I emission lines, which emanate from different excited states of Kr atom, i.e. 586.68, 587.09, and 587.99 nm, as also reported in [39]. That said, it would be impossible to discriminate with this filter the different excited states that could lead to the observed emissions corresponding to the area $A2_{586}$.

In order to verify if other excited states of Kr are produced, emission was recorded using a broader filter (Semrock FF01-747/33-25, centered at 747 nm, 41.1 nm bandwidth). With this filter, the emissions recorded do not emanate from Kr 5p´[3/2]$_2$ and would consist of seven Kr I lines, i.e. 728.73, 742.56, 748.69, 758.74, 760.15, 768.52, and 769.45 nm, exhibiting much larger relative intensities than the three previous lines [39]. In fact, temporal signals captured using this filter were similar to those recorded with the previous one, where a single exponential decay region (i.e. only $A1_{747}$) was always observed for laser energies within the quadratic regime (i.e. $E_{Laser}$≤0.4 μJ/pulse for $P_{Kr}$=3 mbar), while for the saturated regime two distinct areas were produced: $A1_{747}$ over very short timescales, and $A2_{747}$ over long timescales (indicated in **Fig. 3c**, black dashed line at $P_{Kr}$=2 mbar). However, $A2_{747}$ is seen to be much more predominant when compared to $A1_{747}$. These signals became stronger with krypton pressure and laser energy, especially in the saturated regime.

Kr-atom emission that is not exclusively attributed to the relaxation of the fluorescing state Kr 5p´[3/2]$_2$ starts almost immediately after the laser pulse, and lasts for a relatively long time as compared to the lifetime of Kr 5p´[3/2]$_2$ state. Its presence suggests the existence of additional emission mechanisms. These mechanisms are rapidly induced by the laser pulse and remain active over almost 1 μs. It means that these mechanisms are able to produce several emitting excited species over almost 1 μs time-scale.

Furthermore, the presence of the secondary decay region (A2$_{586}$ or A2$_{747}$) for signals recorded in the saturated regime using both interference filters, clearly points towards the existence of a secondary much slower chain of reactions leading to the formation of different excited states of krypton. One of the likely mechanisms which could explain this observation is PIN from Kr 5p´[3/2]$_2$ state through the absorption of a third laser photon [4,5]. This leads to the formation on Kr$^+$ ions, which through transformation to $Kr_2^+$ ions, and subsequent radiative recombination, could produce several Kr excited states [39,40], which de-excite radiatively in the spectral range 420-900 nm [39]. Additional Kr I lines in the TALIF spectrum at high E$_{Laser}$ were also observed in [41] while studying two-photon absorption from Kr atom at 215 nm, and were also attributed to PIN, which further validates our findings. In fact, the energy of Kr 5p´[3/2]$_2$ is of 97945.97 cm$^{-1}$ (12.1436 eV), and the ionization threshold of Kr is of 112914.47 cm$^{-1}$, resulting in an energy gap of 14968.5 cm$^{-1}$ [42]. The laser photon energy in our case is of 49005.19 cm$^{-1}$, which is more than enough to produce Kr$^+$ ions from the 5p´[3/2]$_2$ level. This would in fact explain the departure from the quadratic regime at the higher E$_{Laser}$ studied.

In order to understand the relative significance of the different physical mechanisms involved in the TALIF scheme, a simple 0D model was developed (similar to that used in [24]). For the very short time scales corresponding to the laser pulse, the model takes into account the two-photon excitation rate of the ground-state Kr atom, PIN from Kr 5p´[3/2]$_2$ and Kr 5s[3/2]$_1$ states, and ASE, in addition to the decay of the Kr 5p´[3/2]$_2$ fluorescing state by spontaneous emission (fluorescence) and collisional quenching ($\tau$ values obtained from this work). Ionization from Kr(5p) (cross-section: $\sigma_i \cong 3.6 \times 10^{-18}$ cm$^2$) and Kr(5s) states ($\sigma_i \cong 2.8 \times 10^{-21}$ cm$^2$) [43] leads to the formation of Kr$^+$ ions. Also, ASE results from a population inversion between Kr 5p´[3/2]$_2$ and Kr 5s[3/2]$_1$ states [33,44], leading to an amplification of the fluorescence along the laser beam [45]. ASE has been implemented in the model through the analytical solution given in [46]. PIN and ASE would deplete the Kr 5p´[3/2]$_2$ state leading to saturated fluorescence regimes.

For the large time scales (after the laser pulse), in addition to the $A_3 + Q_3$ loss channels, mechanisms related to the ions Kr$^+$ and $Kr_2^+$ have been considered in the model. The main modes of loss of Kr$^+$ ions are known to be through diffusion or by the formation of $Kr_2^+$ ions, as follows [47,48]:

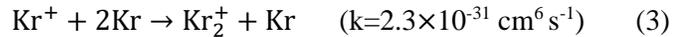

$$Kr^+ + 2Kr \rightarrow Kr_2^+ + Kr \quad (k=2.3\times10^{-31}\ cm^6\,s^{-1}) \quad (3)$$

Then $Kr_2^+$ undergoes dissociative recombination, as follows [39]:

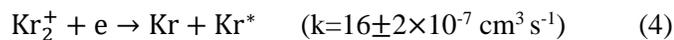

$$Kr_2^+ + e \rightarrow Kr + Kr^* \quad (k=16\pm2\times10^{-7}\ cm^3\,s^{-1}) \quad (4)$$

Here we have assumed that the Kr* excited-state density is equally distributed between Kr 5p´[3/2]$_2$ and Kr 5s[3/2]$_1$ states, which are the only excited states that have been considered in the model.

**Fig. 4** shows the results of the numerical model performed at $P_{Kr}$=3 mbar (as in **Fig. 2**) for the same two $E_{Laser}$ values as in **Fig. 3**. Firstly, the effect of ASE was found to be negligible when compared to PIN even at the higher laser energy. In fact, the intensity of ASE was only significant within the laser pulse (~10 ps), and the depletion of the laser-excited state through ASE was negligible as it accounts for less than 0.1% of the total number of TALIF photons. This is consistent with reference [27], where it was shown experimentally that ASE from H atoms in a $H_2/O_2$ flame was only formed during the laser pulse (35 ps) and for less than 30 ps after it. Thus, the major mode of depletion of the laser-excited state is through PIN.

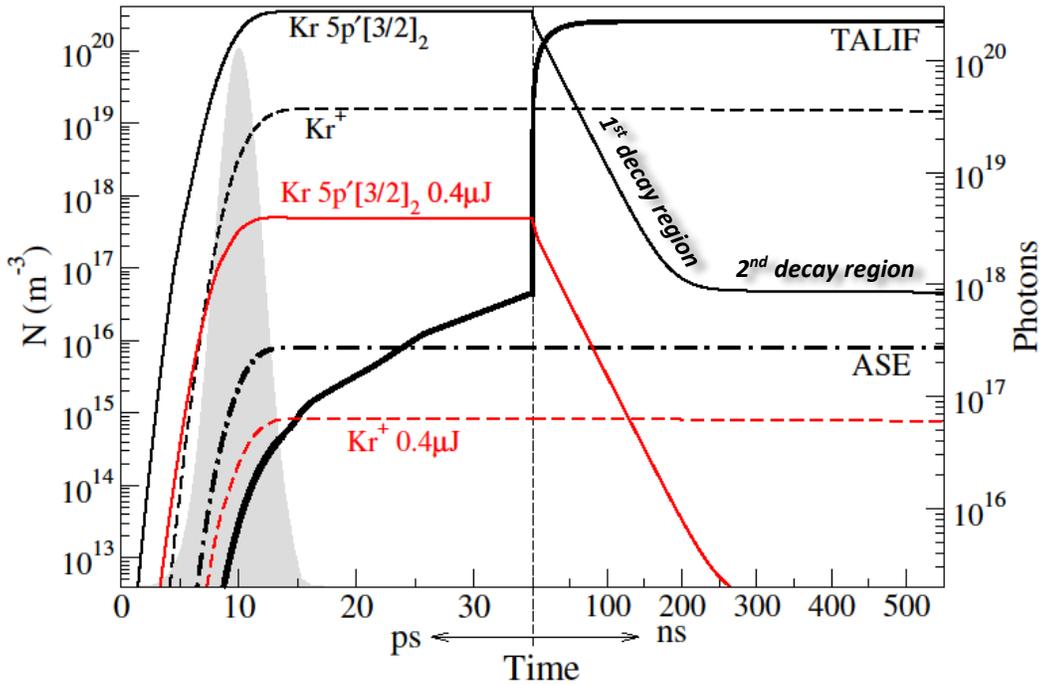

**Figure 4**. Indicative modelling results referring to $P_{Kr}$=3 mbar. Black and red solid lines are the Kr 5p´[3/2]$_2$ densities (N) obtained for $E_{Laser}$=12.1 and 0.4 μJ/pulse, respectively. Also shown for $E_{Laser}$=12.1 μJ/pulse are: i) the cumulative number of photons corresponding to ASE (dash-dot) and TALIF (thick black), and ii) the density (N) of Kr$^+$ ions at both $E_{Laser}$ values (dash). The shaded grey curve represents the laser pulse.

The results obtained by the model may be better understood by a simple rate analysis. As a matter of fact, using the available experimental information, the PIN rate ($\Gamma$) of Kr(5p) states can be qualitatively estimated as follows [27]:

$$\Gamma = \frac{\sigma_i I}{hc\tilde{\nu}_{c_{Kr}}} \qquad (5)$$

For $I=I_{max}$=484 MW.cm$^{-2}$, eq. (5) gives $\Gamma$=1.79×10$^9$ s$^{-1}$. This is almost two orders of magnitude larger than $A_3 + Q_3$, which herein is estimated to be of ~4.25×10$^7$ s$^{-1}$ at $P_{Kr}$=3 mbar, and ~6.5×10$^7$ s$^{-1}$ at $P_{Kr}$=10 mbar, in good agreement with references [7,37,38]. Furthermore, for $I$=18 MW.cm$^{-2}$, i.e. the upper intensity limit of the quadratic regime in **Fig. 2**, $\Gamma$=6.67×10$^7$ s$^{-1}$, which is still slightly higher than $A_3 + Q_3$ at 3 and 10 mbar. Therefore, similarly to what is suggested by the model, PIN determines the population of the fluorescing state and, thus, the TALIF signal intensity beyond the quadratic regime. On the other hand, for $I$=4 MW.cm$^{-2}$ ($E_{Laser}$=0.1 µJ/pulse, i.e. the lowest $E_{Laser}$ giving an acceptable TALIF signal-to-noise ratio in our work), $\Gamma$ decreases to 1.48×10$^7$ s$^{-1}$. This is ~65% and ~77% lower than $A_3 + Q_3$ at 3 and 10 mbar, respectively. This result is also supported by **Fig. 4**, where the density of Kr$^+$ ions at $E_{Laser}$=0.4 µJ/pulse (dashed red) is 4 orders of magnitude lower than that at $E_{Laser}$=12.1 µJ/pulse (dashed black). Thus, PIN and ASE effects are limited in the quadratic regime, fulfilling the condition (*c2*) above.

Further, for the higher $E_{Laser}$ in **Fig. 4**, the model captures the transition from a 1$^{st}$ to a 2$^{nd}$ signal-decay region (both indicated in the figure), which starts at the inflection point around 200 ns. This transition is similar to that captured experimentally in **Fig. 3b** ($P_{Kr}$=3 mbar), the corresponding signal-decay regions denoted by A1$_{586}$ and A2$_{586}$, respectively. This establishes that PIN is the major saturation mechanism of the TALIF signal due the creation of Kr$^+$ ions, followed by reactions (3) and (4), and radiative relaxations of Kr* (formed through reaction (4)). In fact, the slope of $S_{\tilde{v}t}$ versus $E^2_{Laser}$ in the saturation regime is of ~0.5 (**Fig. 2**), which also strongly suggests that depletion of Kr 5p´[3/2]$_2$ takes place through PIN. Also to be noted is that the contribution of the 2$^{nd}$ decay region to the fluorescence yield is negligible.

The key-contribution of PIN to the TALIF signal properties may be also inferred from the spectral analysis of the two-photon absorption line (**Fig. 5**).

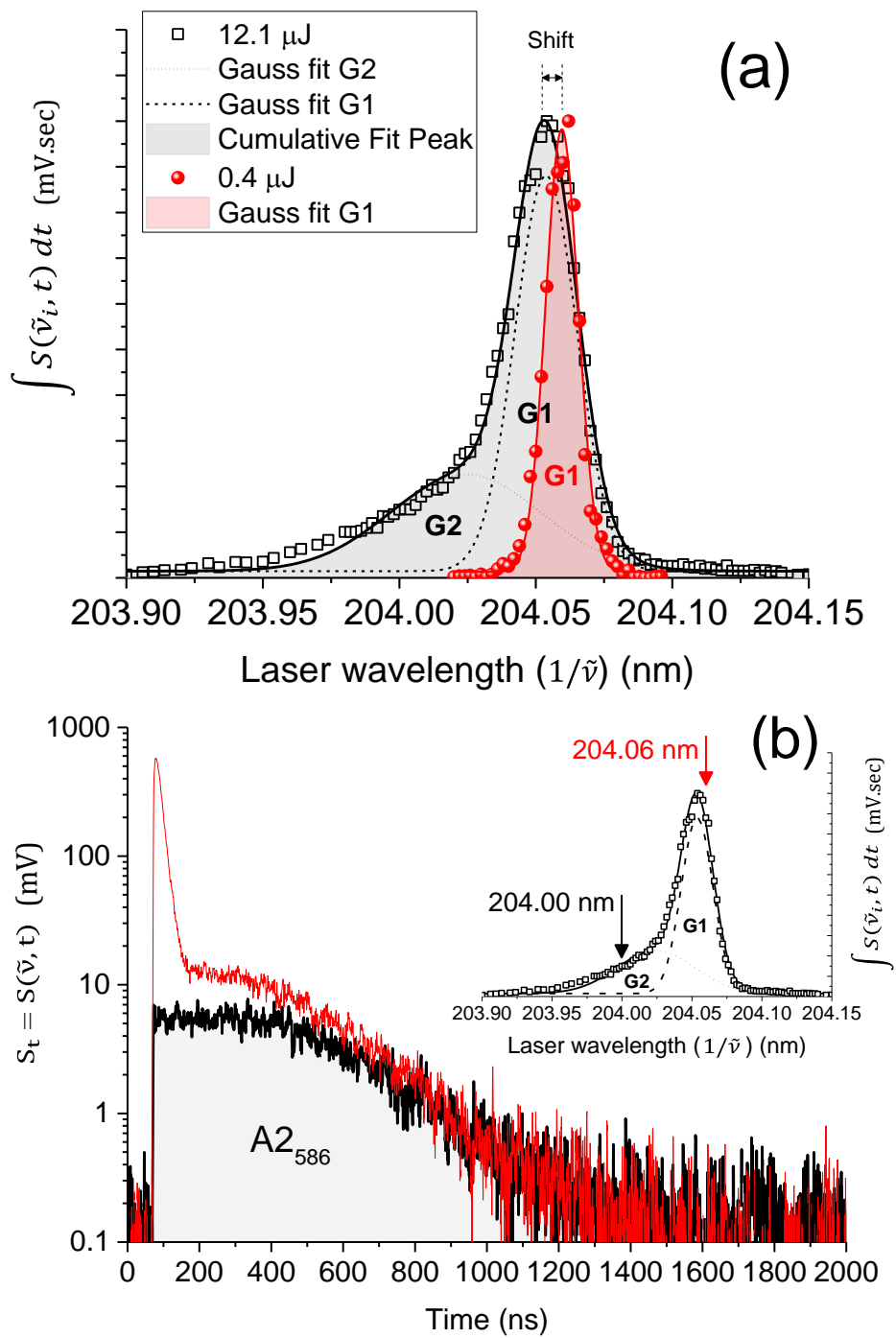

**Figure 5.** (a) Experimental high-resolution two-photon absorption line profiles (symbols), $\int S(\tilde{v}_i, t)\, dt$, fitted with Gaussians (solid/dashed lines) for low (0.4 μJ/pulse – red) and high (12.1 μJ/pulse – black) $E_{Laser}$, respectively, at $P_{Kr}$=3 mbar. At the higher $E_{Laser}$, the center of the absorption line is shifted by –6 pm. (b) PMT signals $S_t=S(\tilde{v},t)$ at $P_{Kr}$=3 mbar and $E_{Laser}$=12.1 μJ/pulse, recorded with a 586 nm-centred interference filter at two laser wavelengths (indicated by the arrows on the corresponding cumulative spectral profile in the inset): 204.00 nm (black) and 204.06 nm (red).

13**Fig. 5a** illustrates two normalized spectral profiles obtained at $P_{Kr}$=3 mbar for the lower- and upper-limit values of $E_{Laser}$ in **Figs. 3** and **4**. To the best of our knowledge, such a detailed analysis of the two-photon absorption line profile, especially for $E_{Laser}$ values outside the quadratic regime, has not been performed in the literature regarding ultrafast TALIF studies in Kr. The experimental data points (black squares and red spheres) were obtained through the temporal integration of $S(\tilde{v}_i, t)$, which was recorded at different laser wavelengths around the central laser frequency. For $E_{Laser}$=0.4 μJ/pulse, the spectral profile (red spheres) is narrow and can be well-fitted with one Gaussian (red line, denoted as G1). In this case, the main broadening mechanism of the two-photon absorption line is the laser instrumental function, which has a Gaussian shape. As a matter of fact, the only other broadening effect that can enter into play is the Doppler broadening for Kr atoms that turns out to be much smaller than the laser linewidth (0.14 versus ∼17 pm, respectively). For $E_{Laser}$=12.1 μJ/pulse, however, the recorded spectral profile (black open squares) can be well-fitted only by combining two Gaussians (denoted as G1 and G2 in **Fig. 5a**). While the peak G1 corresponds to a two-photon excitation to the Kr fluorescing state of interest (i.e. $4p^6$ $^1S_0 \rightarrow \rightarrow$ Kr 5p´[3/2]$_2$ transition), as expected, the appearance of the peak G2 strongly indicates a contribution from other processes. In fact, as it can be seen in **Fig. 5b**, the formation of the peak G2 is directly associated to the appearance of the emissive region A2$_{586}$ in the time-range ∼80–1000 ns of the PMT signal. Indeed, when the laser is tuned to 204.00 nm (black signal, see corresponding black arrow in the inset spectral profile), only emission from region A2$_{586}$ is recorded. Furthermore, when tuning the laser at the peak two-photon absorption frequency (red signal/arrow), A2$_{586}$ is still present along with the fluorescence peak, which is seen between ∼80 and 200 ns. Thus, at the higher $E_{Laser}$, the existence of A2$_{586}$ and G2 indicates the presence of additional emission channels that are initiated by the laser, they last over almost 1 μs, and they are captured when using this filter. PIN is a process that may explain this behavior, as follows. First, it initiates the slow chain of reactions (3) and (4), as discussed through the analysis of **Figs. 3** and **4**, which would result to the creation of different Kr* states after the laser pulse. Therefore, these would contribute to the formation of the long-lasting part of the region A2$_{586}$ through spontaneous emissions [39-41,51]. Second, the formation of A2$_{586}$ in **Fig. 5b** happens very fast, almost immediately after the laser pulse. This may be also attributed to PIN through the generation of photo-electrons. Depending on their energy and density, these would rapidly enter into collisions with Kr atoms, leading to their fast excitation, especially if the photo-electrons are substantially heated by the laser electric field and the ionization degree is high. This seems to be the case at the higher $E_{Laser}$ studied herein, as revealed by the model (**Fig. 4**).

Putting it all together, a likely sequence of processes that would explain the obtained results at the higher $E_{Laser}$ is the following: (i) the application of the laser pulse leads to two-photon excitation of Kr at the fluorescing state Kr 5p´[3/2]$_2$, (ii) laser-induced photoionization of this state happens very fast (within the laser pulse), leading to the production of photo-electrons; their energy is estimated to be of ∼4.2 eV in our case (based on the difference between the laser photon energy, and the energy gap between the Kr ionization threshold and the Kr 5p´[3/2]$_2$ state energy, as discussed above), (iii) generated photo-electrons may be further heated by the laser electric field, inducing a fast excitation of Kr (depending on their energy and density), and a subsequent early emission, which is seen in the very beginning of the PMT signal in **Fig. 5b**, (iv) the moderately-fast reaction (3) leads to the formation of $Kr_2^+$ ions, which through radiative recombination (reaction (4) – slow process) lead to the creation of Kr* states, (v) radiative relaxation of Kr* states induces the long-lasting emission part of the region A2$_{586}$ in **Figs. 3** and **5b**.

However, in spite of having the peak G2 close to the main excitation peak G1, this does not prevent using G1 for calibration purposes when aiming on the measurement of H- and N-atom densities in plasmas. This is due to the negligible contribution from G2 when using the 586-centred filter for $E_{Laser}$ values in the quadratic regime, as seen in **Fig. 5a** for $E_{Laser}$=0.4 µJ/pulse (red signal). Besides, the area A2$_{586}$ in the time variation of the PMT signal also disappears for $E_{Laser}$≤0.4 µJ/pulse (see **Figs. 3b** and **4**, red signals).

Furthermore, in **Fig. 5a**, we also observe that at the higher $E_{Laser}$ the line G1 is noticeably broader, and its central wavelength is shifted by –6 pm from the central laser wavelength. This should be attributed to an AC Stark shift induced by the linearly-polarized electromagnetic field of the laser [26,27,44,49]. Besides, the possible mechanism for the observed broadening may be the Stark effect due to the presence of charged species [44,52]. Indeed, the results obtained from the investigation of the fluorescence time-decay, the laser-induced emission, the spectral analysis of the two-photon absorption line, and the laser-absorption-induced collisional-radiative kinetics, give a strong indication for the presence of charged species, which create a strong local micro-field resulting in a Stark effect. This induces the two-photon absorption-line broadening and shift. Our results would indicate that this effect would quickly dominate the laser absorption mechanism and result in the deviation from the quadratic regime when $E_{Laser}$ is increased. In fact, this mechanism has been also proposed in [44], where the excitation line profiles corresponding to the two-photon excitation of ground-state Kr atom to the 5p[5/2]$_2$ state, and its subsequent transitions to the Kr 5s[3/2]$_1$ or Kr 5s[3/2]$_2$ states, were broader than the laser line, similar to our case. It should be noted that broadening due to power-saturation [50] of the optical transition is ruled out for our conditions since $W_{ij} \times \tau_L$<<1. Besides, the power saturation broadening always comes with a much less increase of the peak fluorescence with $E_{Laser}$ (not seen in **Figs. 3** and **4**).

**Conclusions**

In summary, ps-TALIF in Kr between 0.1–10 mbar was fundamentally investigated in this work. The laser intensity (*I*) was varied between 1 and 480 MW.cm$^{-2}$. The influence of photoionization (PIN) and amplified stimulated emission (ASE) on the TALIF signal properties was detailly studied for a Kr gas pressure of 3 mbar. PIN was the dominant process leading to the depletion of the laser-excited state (5p´[3/2]$_2$), which became increasingly significant for $15 < I \leq 480$ MW.cm$^{-2}$. Thus, a saturated fluorescence regime was obtained, and broader two-photon absorption-line profiles probably due to the Stark effect. However, for $I \leq 15$ MW.cm$^{-2}$, the PIN effect was significantly limited, the two-photon absorption line was narrow, and the quadratic dependence of the TALIF-signal intensity versus the laser energy was revealed. Thus, in this case the investigated Kr TALIF scheme, including two-photon excitation to the 5p´[3/2]$_2$ state and fluorescence at 587.09 nm can be used for calibration purposes in ps-TALIF studies. Besides its fundamental interest, this work contributes to the development of ps-TALIF diagnostics for measuring absolute densities and quenching coefficients of H and N atoms in reactive plasmas and/or flames.

## Acknowlegments

This work was supported by the French 'Agence Nationale de la Recherche' (ANR) through the ASPEN project (grant ANR-16-CE30-0004/ASPEN), and by the SESAME research and innovation programme of the Ile-de-France Region under the project grant DIAGPLAS.